\documentclass{article}
\usepackage{graphicx}
\usepackage{subfigure}
\usepackage{relsize}
\usepackage{amssymb}
\usepackage{xfrac}
\usepackage{amsmath}
\usepackage{amsfonts}
\usepackage{epsfig}
\usepackage{bm}
\usepackage{multirow}

\newcommand{\bp}{\mbox{\boldmath $p$}}

\newcommand{\bk}{\mbox{\boldmath $k$}}

\graphicspath{{graphics/}}

\begin{document}

\title{ Exclusive $\pi^+\pi^-$ Production at the LHC\\ with Forward Proton Tagging }

\newcommand{\comment}[1]{$^*$\marginpar{\bf $^*$#1}}

\author{
R. Staszewski$^a$, P. Lebiedowicz$^a$, M. Trzebi\'nski$^a$,\\J. Chwastowski$^{b,a}$ and A. Szczurek$^{a,c}$ \\[7mm]
\multicolumn{1}{p{.99\textwidth}}{\smaller{ { $^a$
Institute of Nuclear Physics Polish Academy of Sciences,\newline
ul. Radzikowskiego 152, 31-342 Krak\'ow, Poland. }}}\\[7mm]
\multicolumn{1}{p{.99\textwidth}}{\smaller{ { $^b$
Institute of Teleinformatics\newline
Faculty of Physics, Mathematics and Computer Science,\newline
Cracow University of Technology,\newline
ul. Warszawska 24, 31-115 Krak\'ow, Poland.  }}}\\[15mm]
 \multicolumn{1}{p{.99\textwidth}}{\smaller{ { $^c$
University of Rzesz\'ow\newline
Aleja Rejtana 16c, 35-959 Rzesz\'ow, Poland. }}}
}

\maketitle

\begin{abstract}
A process of Central Exclusive $\pi^+\pi^-$ production in proton-proton
collisions and its theoretical description is presented. A possibility of its
measurement, during the special low luminosity LHC runs, with the help of the ATLAS
central detector for measuring pions and the ALFA stations for tagging the
scattered protons is studied. A visible cross section is estimated to be 
21~$\mu$b for $\sqrt{s}=7$~TeV, which gives over 2000 events for 100~$\mu$b$^{-1}$ of
integrated luminosity.  Differential distributions in pion pseudorapidities, pion and
proton transverse momenta as well as $\pi^+\pi^-$ invariant mass are shown and
discussed.
\end{abstract}

\section{Introduction}

Processes of central exclusive production have gained a lot of interest in the
recent years. Although the attention is paid mainly to high-$p_T$ processes
that can be used for new physics searches (exclusive Higgs, $\gamma\gamma$
interactions, \textit{etc.}), measurements of low-$p_T$ signals are also very
important as they can help to constrain models and to understand the
backgrounds for the former ones. The aim of this paper is to discuss a
measurement possibility of the exclusive $\pi^+\pi^-$ production at the LHC, in
which both protons are detected. 

This reaction is a natural background for exclusive production of resonances
decaying into the $\pi^+\pi^-$ channel, such as: $f_2(1270)$, glueballs,
glueball candidates (\textit{e.g.} $f_0(1500)$) or charmonia (\textit{e.g.}
$\chi_c(0)$). Since the cross section for the $pp\rightarrow p\pi^+\pi^- p$
process is fairly large \cite{LS10} and estimation is important for the exotic
meson production. 

The dominant mechanism  of the $pp\rightarrow p\pi^+\pi^- p$ reaction is
relatively simple compared to that of the $pp\rightarrow n\pi^+\pi^+ n$
\cite{pp_to_nnpi0pi0} or $pp\rightarrow p\pi^0\pi^0 p$ processes. However, the
only measurements at high energies were performed at the CERN ISR for
$\sqrt{s}=62$~GeV \cite{ABCDHW89, ABCDHW90} and $\sqrt{s}=63$~GeV
\cite{AFS}. A measurement at higher energy can add to the understanding of the
diffractive reaction mechanism.

\section{Theoretical Model}
\label{section:Theoretical_Description}

\begin{figure}[!ht]
\centering
\includegraphics[width=0.35\textwidth]{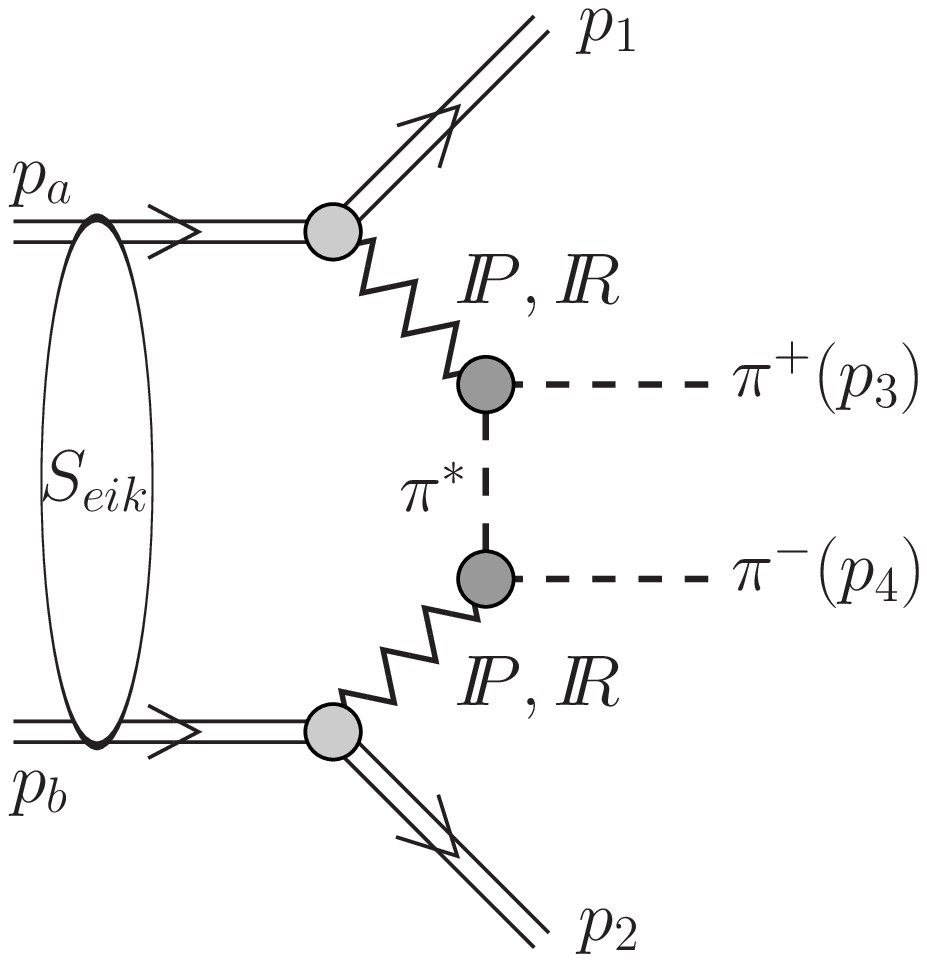}
\hspace{0.1\textwidth}
\includegraphics[width=0.35\textwidth]{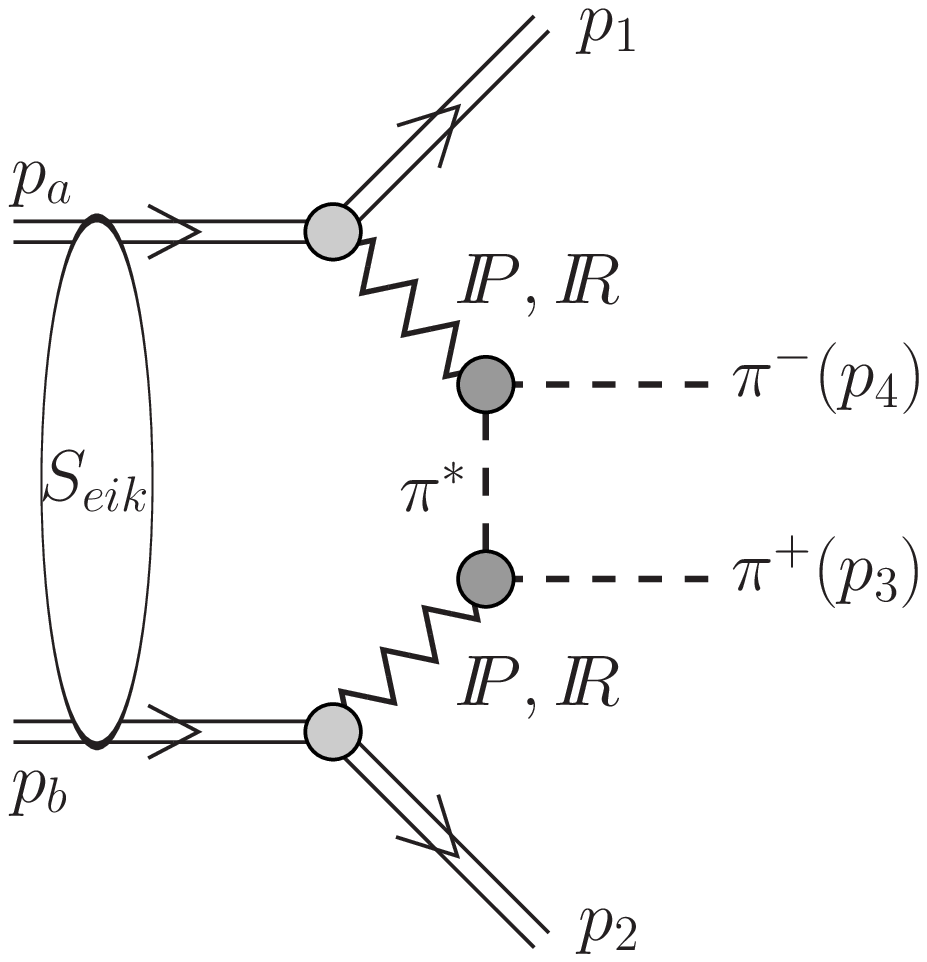}
   \caption{\label{fig:central_double_diffraction_diagrams}
The double-diffractive mechanism of exclusive production of
$\pi^{+}\pi^{-}$ pairs including the absorptive corrections. }
\end{figure}

The dominant mechanism of the exclusive production of the $\pi^{+}\pi^{-}$ pairs at
high energies is sketched in
Fig.~\ref{fig:central_double_diffraction_diagrams}.  The formalism used in
calculations is explained in detail elsewhere \cite{LS10,LPS11} and here only
the main aspects are discussed.

The full amplitude for the exclusive process $pp \rightarrow p \pi^+ \pi^- p$
(with four-momenta $p_{a}+p_{b}\rightarrow p_{1}+p_{3}+p_{4}+p_{2}$) is a sum
of the bare and rescattering amplitudes:
\begin{equation} {\cal {M}}^{full} = {\cal {M}}^{bare} + {\cal {M}}^{rescatt}
  \,.  \label{amp_pipi_full} \end{equation}
The bare amplitude can be written as:
\begin{eqnarray} \nonumber \mathcal{M}^{bare}&=&
  M_{13}(s_{13},t_1)F_{\pi}(t_{a})\frac{1}{t_{a}-m_{\pi}^{2}}F_{\pi}(t_{a})M_{24}(s_{24},t_2)\\
  &+&M_{14}(s_{14},t_1)F_{\pi}(t_{b})\frac{1}{t_{b}-m_{\pi}^{2}}F_{\pi}(t_{b})M_{23}(s_{23},t_2)\;
  , \label{Regge_amplitude} \end{eqnarray}
where $M_{ik}$ denotes the coupling between: forward proton ($i=1$) or backward
proton ($i=2$) and one of the two pions ($k=3$ for $\pi^{+}$, $k=4$ for
$\pi^{-}$). The energy dependence of the $\pi p$ elastic amplitudes
is parameterised in terms of Regge theory by Pomeron and Reggeon exchanges. The
values of coupling constants  and the Regge trajectory parameters are taken
from the Donnachie-Landshoff analysis of the total and elastic cross sections
for $\pi N$ scattering \cite{DL92}. The slope parameters of the elastic $\pi p$
scattering are taken as 
\begin{equation} B(s) = B_{i} + 2 \alpha'_{i} \ln \left( \sfrac{s}{s_0}
  \right), \end{equation}
where $B_{I\!\!P}$ = 5.5 GeV$^{-2}$, $\alpha'_{I\!\!P}$ = 0.25 GeV$^{-2}$ and
$B_{I\!\!R}$ = 4 GeV$^{-2}$, $\alpha'_{I\!\!R}$ = 0.93 GeV$^{-2},$ for Pomeron
and Reggeon exchanges, respectively and $s_{0}$ = 1 GeV$^{2}$.

The Donnachie-Landshoff parametrisation is used only for the $\pi p$ subsystem
energy $W_{ik}>2$ GeV (see Ref.~\cite{LS10}). Below this energy resonance states
are present in $\pi p$ subsystems in single diffractive processes.  Their
contribution would appear at large rapidities, so it is not discussed here.  In
order to exclude resonance regions the $M_{ik}$ terms are corrected by  purely
phenomenological smooth cut-off correction factors which modify the cross section
only at large rapidities \cite{LS10}.

The form factors, $F_{\pi}(t)$, correct for the off-shellness of the
intermediate pions in the middle of the diagrams shown in
Fig.~\ref{fig:central_double_diffraction_diagrams}.  In the following they are
parameterised as:
\begin{equation}
  F_{\pi}(t)=\exp\left(\frac{t-m_{\pi}^{2}}{\Lambda^{2}_{off}}\right)\,,
  \label{off-shell_form_factors} \end{equation}
where the parameter $\Lambda_{off}$ is not known precisely.  In~\cite{LPS11} a
fit to the experimental data \cite{ABCDHW90} was performed and
$\Lambda_{off}^2=2$~GeV$^2$ was obtained. This value was used in the
calculations for $pp\rightarrow p\pi^+\pi^- p$ production at $\sqrt{s}=7$~TeV
and resulted the cross section of 234~$\mu$b.

The absorptive corrections to the bare amplitude, marked in
Fig.~\ref{fig:central_double_diffraction_diagrams} by the blob, were taken into
account in \cite{LPS11} as:
\begin{equation} {\cal M}^{rescat}= \mathrm{i} \int \frac{d^{2}
  \bk_{t}}{2(2\pi)^{2}} \frac{A_{pp}(s,k_{t}^{2})}{s} {\cal M}^{bare}
  (\bp^{\,*}_{a,t}-\bp_{1,t},\bp^{\,*}_{b,t}-\bp_{2,t}) \;,
  \label{abs_correction} \end{equation}
where $\bp^{\,*}_{a} = \bp_{a} - \bk_{t}$, $\bp^{\,*}_{b} = \bp_{b} + \bk_{t}$
and $\bk_t$ is the transverse momentum exchanged in the blob.  The amplitude
for elastic proton-proton scattering is parameterised as:
\begin{equation} A_{pp}(s,k_{t}^{2})= A_{0}(s) \, \exp(-B k_{t}^{2} /2)\, .
  \label{pp_scatt} \end{equation}
The optical theorem gives: Im$A_{0}(s,t = 0) = s \sigma_{tot}(s)$ and the real
part is small in the high energy limit.  The Donnachie-Landshoff
parametrisation of the total and elastic $pp$ or $p\bar{p}$ cross sections is
used to calculate the rescattering amplitude and $B^{pp}_{I\!\!P}$ = 9
GeV$^{-2}$ value is taken.

The cross section is obtained by integration over the four-body phase space,
which was reduced to 8 dimensions and  performed numerically. A weighted Monte
Carlo generator based on this model has been developed and was used in the following
analysis.

\section{Experimental Setup}

\begin{figure}[ht]
  \centering
  \includegraphics[width=0.9\textwidth]{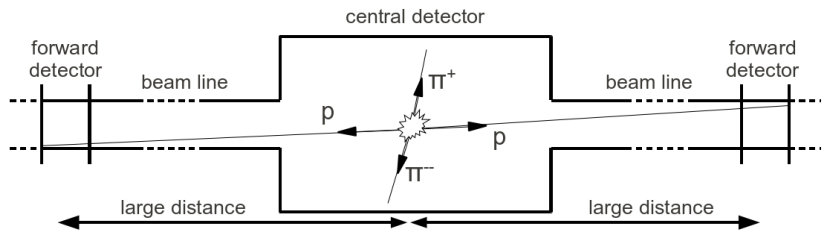}
  \caption{A scheme of the measurement concept -- pions are registered in the
  central detectors, whereas protons in the very forward
  detectors.}
  \label{fig:scheme}
\end{figure}

The final state of the $pp\rightarrow p\pi^+\pi^- p$ consists of four particles
-- two protons and two pions. At the LHC the pions are produced in the rapidity
range $|y|<10$, whereas the protons are scattered at very small angles (of the
order of microradians) into the accelerator beam pipe.  Therefore, to perform a
fully exclusive measurement, there is a need of two different types of detectors (see
Fig.~\ref{fig:scheme}): a central detector (for pion detection) and very
forward detectors (for proton tagging). The analysis presented in this paper
assumes ATLAS as the central detector and ALFA as the proton tagging detectors.

The ATLAS detector \cite{ATLAS} is located at the LHC Interaction Point 1
(IP1).  It has been designed as a general purpose detector with a large
acceptance in pseudorapidity, full azimuthal angle coverage, good charged
particle momentum resolution and a good electromagnetic calorimetry completed
by full-coverage hadronic calorimetry. The ATLAS tracking detector provides
measurement of charged particles momenta in the $|\eta| < 2.5$ region and the
calorimeter covers $|\eta|<4.9$.

The ALFA (Absolute Luminosity For ATLAS) detectors \cite{TDR} are designed for
proton-proton elastic scattering measurement in the Coulomb-nuclear amplitude
interference region. These detectors are placed about 240~m from the IP1,
symmetrically on both sides, inside roman pots. These are special devices that allow to
place detectors inside the beam pipe and to control the distance between their
edge and the proton beam. This is of primordial importance for the detectors
safety, since the proton beam can cause serious radiation damage.  Measurement of
protons scattered at very small angles (like in the elastic scattering)
requires a special tune of the LHC accelerator with very small angular dispersion
at the IP. This is granted by a high value of the betatron function
($\beta^*$).  Due to limited radiation hardness, the ALFA detector will be used
only during the dedicated runs.

It is worth mentioning that at the LHC, apart from ALFA,  there are also
similar stations of the TOTEM \cite{TOTEM} experiment placed around the CMS
central detector. Although, the present study was carried out for ALFA and
ATLAS, similar results can be expected for TOTEM and CMS. In addition, two
other proton tagging detectors are presently at the planning stage -- AFP
(ATLAS Forward Proton) for ATLAS and HPS (High Precision Spectrometer) for CMS.
Their purpose is to tag forward protons during high luminosity LHC runs and to look at
high-$p_T$ signals.  The acceptance of these detectors will be completely
different than the one of ALFA and TOTEM.  Actually, the AFP and HPS detectors
will be able to detect protons which lost some part of their initial energy
\cite{Staszewski:2009sw} and will not register protons originating from elastic
scattering. Since for the $pp\rightarrow p\pi^+\pi^- p$ process the energy loss
of the protons is rather small, only the tails of this signal could be seen in
AFP or HPS. Taking into account the fact that these detectors will work during
normal LHC runs, when there will be many independent interactions in one bunch
crossing, it is clear that exclusive pion pair production can be measured only
with help of ALFA or TOTEM.

\section{Results of the Simulation}

A crucial element of the $pp\rightarrow p\pi^+\pi^-p$ measurement is the
tagging of the forward protons with the ALFA detectors. Thus, a very important
ingredient of this analysis is a proper simulation of the proton transport from
the Interaction Point to the ALFA stations through the LHC magnetic lattice.
One needs to remember that the ALFA detectors are designed only for the special
LHC runs so a corresponding description of the LHC magnets has to be used in
the simulation.  In this paper the $\sqrt{s} =7$~TeV and $\beta^*=90$~m LHC
optics was taken, since this is the configuration planned for the ALFA runs in
2011 \cite{chamonix_shutdown}.

\begin{figure}[ht]
  \includegraphics[width=0.49\textwidth]{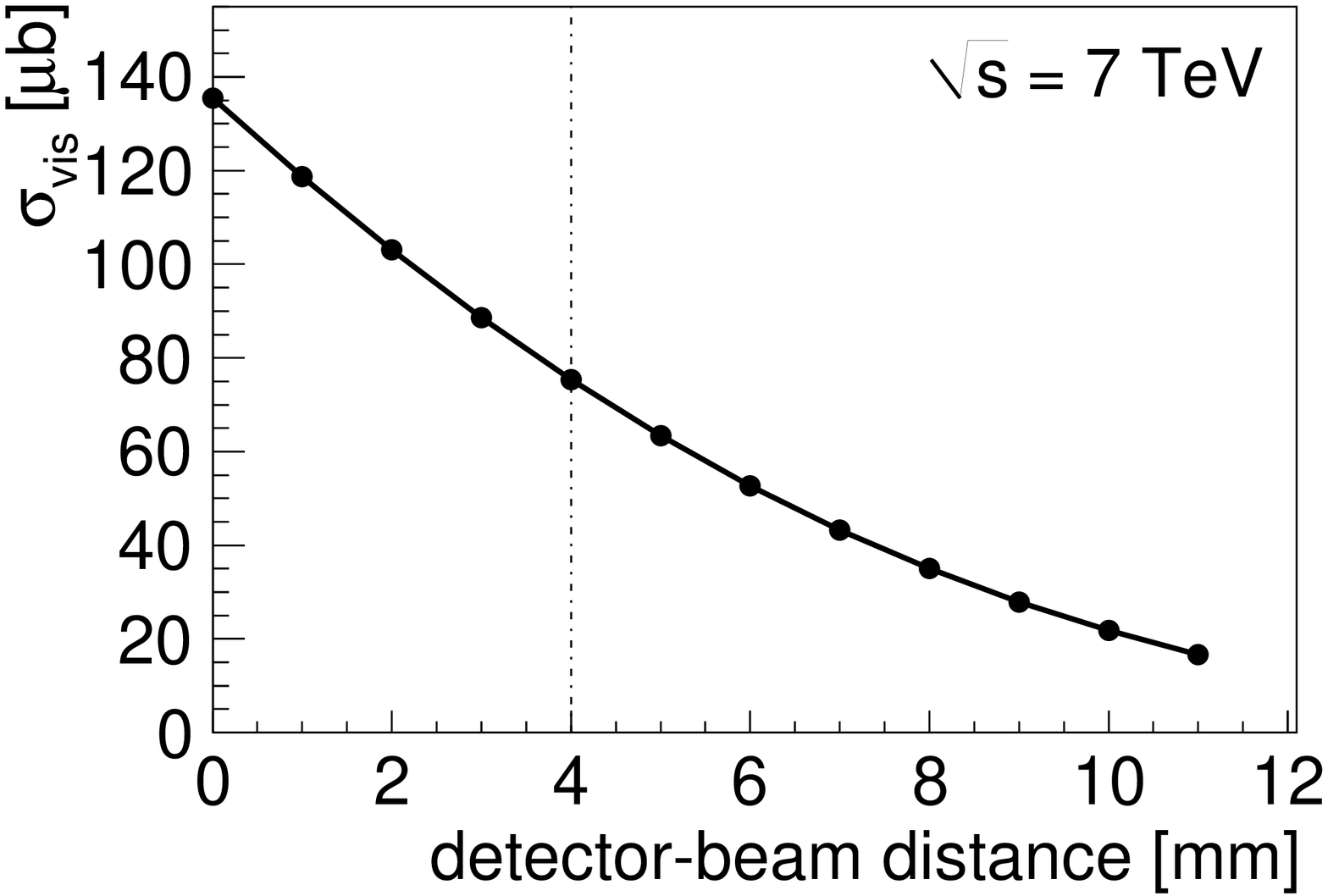}
  \hfill
  \includegraphics[width=0.49\textwidth]{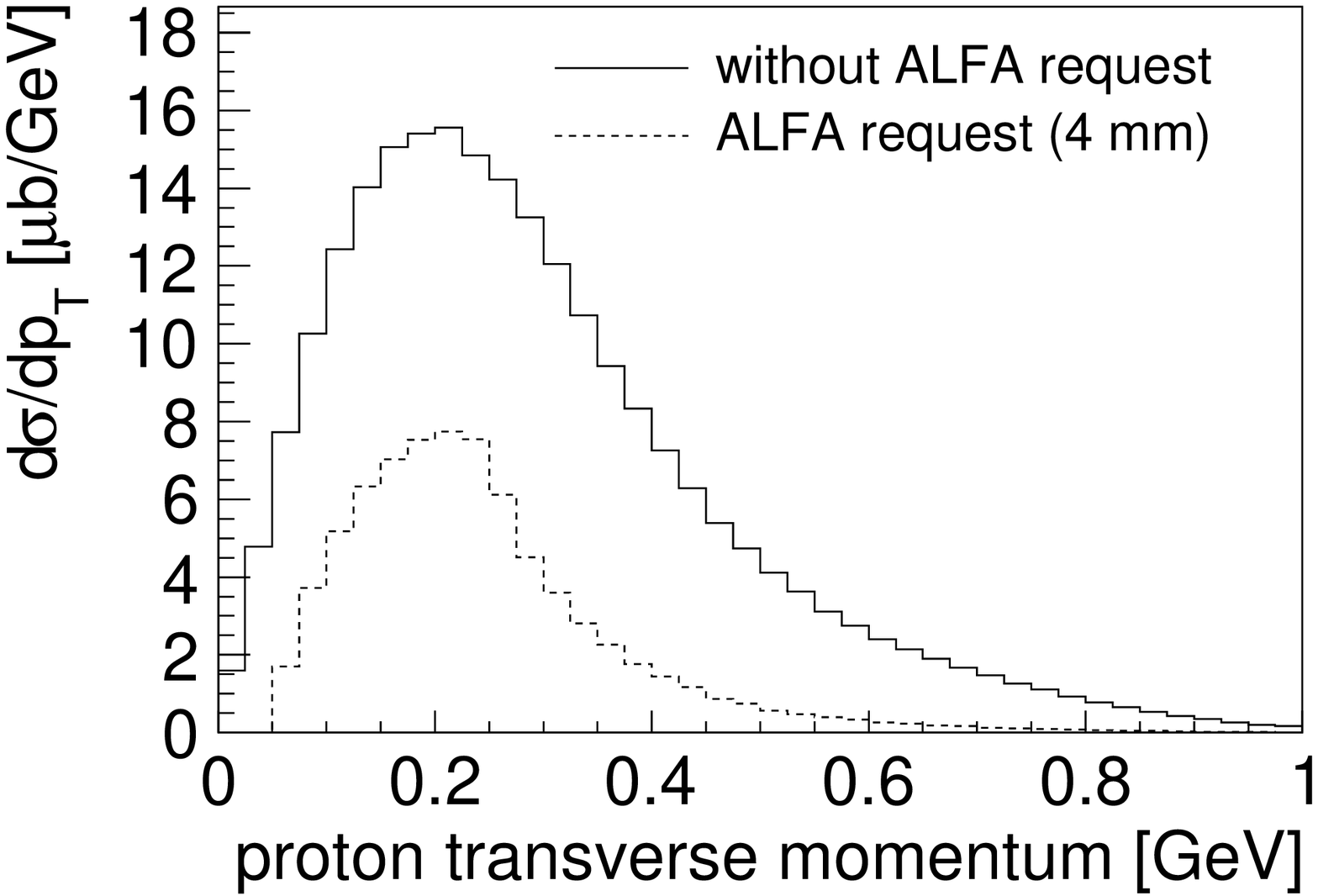}
  \caption{Left: the cross section for $pp\to p\pi^+\pi^- p$ with both protons
  tagged by the ALFA detectors as a function of the distance between the
  detectors edge and the beam centre (assumed to be identical in all ALFA
  stations).  Right: the proton transverse momentum distribution; the dotted
  line marks the distribution for the events with both protons tagged by ALFA
  detectors positioned at 4~mm.}
  \label{fig:alfa_effect}
\end{figure}

The cross section for exclusive $\pi^+\pi^-$ production at $\sqrt{s}=7$~TeV is
234~$\mu$b (see Sec. 2). The requirement that both protons are tagged in the
ALFA stations causes that not all events can be fully registered.  This is due
to limited acceptance of the detectors. In fact, the visible cross section
depends on the distance between the ALFA detector edge and the beam centre (it
will be changed during runs, accordingly to beam conditions).  This dependence
is presented in Fig.~\ref{fig:alfa_effect}~(left). For the rest of the present
analysis a distance of 4~mm is assumed, which corresponds to 75~$\mu$b of cross
section visible in the ALFA detectors. Fig.~\ref{fig:alfa_effect} (right)
presents the distribution of forward proton transverse momentum before and
after requesting that protons are tagged in ALFA.

\begin{figure}[ht]
  \includegraphics[width=0.49\textwidth]{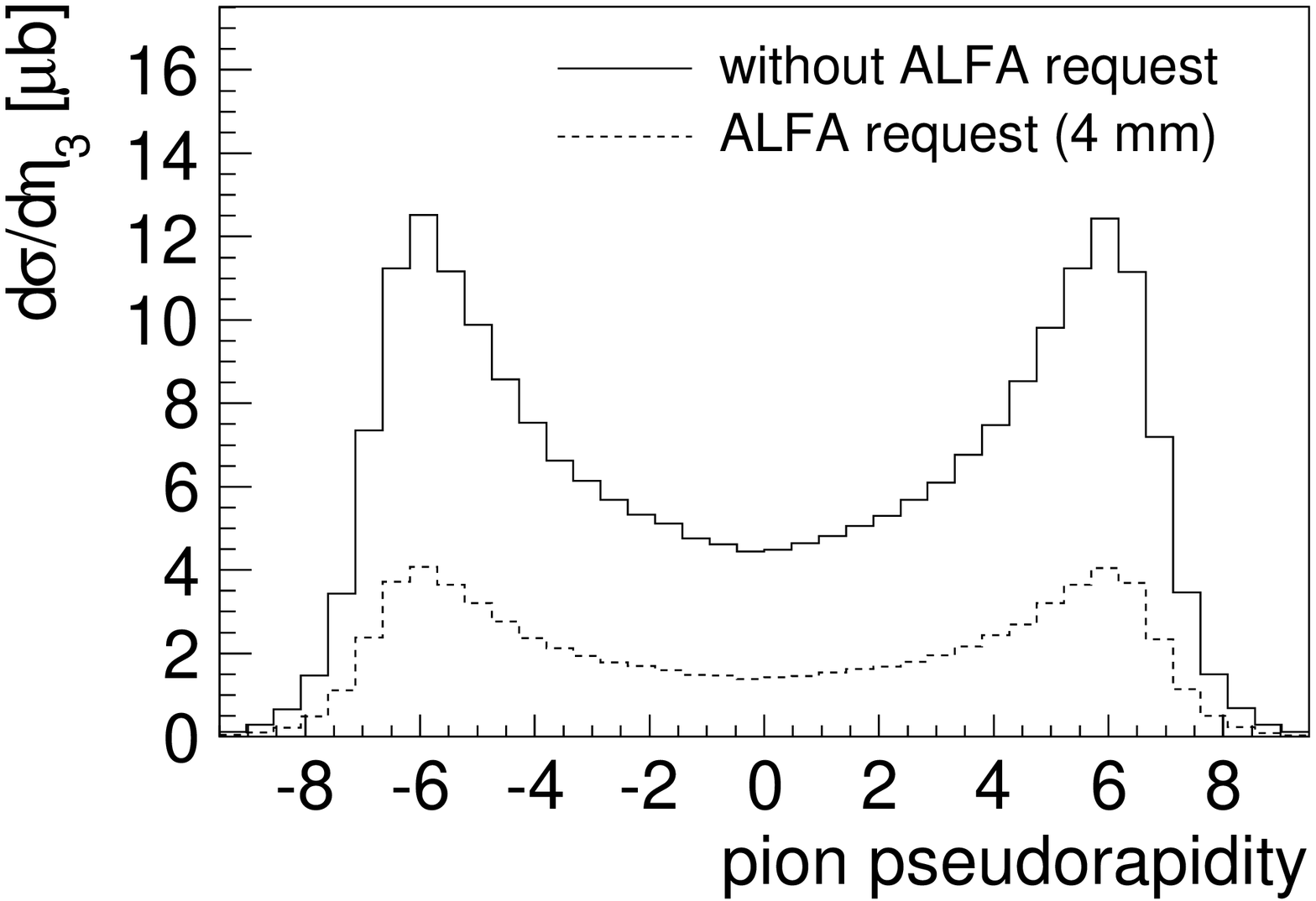}
  \hfill
  \includegraphics[width=0.49\textwidth]{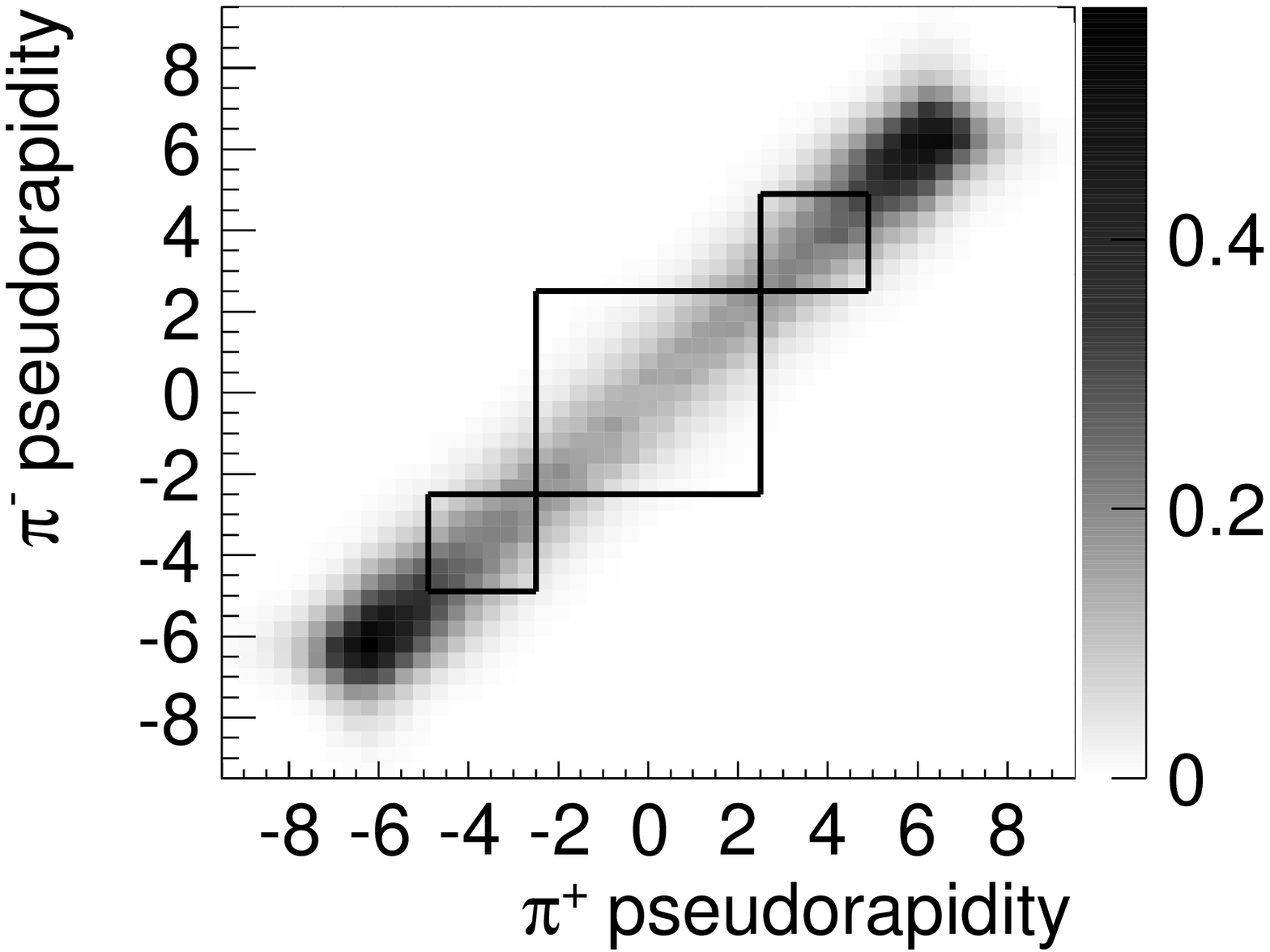}
  \caption{Left: the total cross section as a function of pion pseudorapidity.
  Right: the correlation between the pseudorapidies of the, black frames represent
  regions of tracker and forward calorimeters.}
  \label{fig:eta_dist}
\end{figure}

Pions produced in the discussed process will be measured in the central
detector. The pion pseudorapidity distribution is presented in
Fig.~\ref{fig:eta_dist}~(left), whereas the right panel of
Fig.~\ref{fig:eta_dist} shows a correlation between pseudorapidities of both
pions. The model used for the simulation predicts a strong correlation between
the pseudorapidities of $\pi^+$ and $\pi^-$, which is not expected for pions
originating from $pp\to p\pi^+\pi^0\pi^- p$ or $pp\to p\pi^+\pi^-\pi^+\pi^- p$
processes\footnote{Such reactions are a natural background, when only two pions
are inside the detector acceptance. This contribution can be estimated
experimentally by studying the three and four pion final states.}. Although the
majority of the events contains pions with $\eta$ too large to be detected in
ATLAS, the remaining cross section is still large enough to make the
measurement possible.

The pions can be detected in the ATLAS tracking detector ($|\eta|<2.5$) or in
the ATLAS calorimetry system ($|\eta|<4.9$).  From the experimental point of
view these are two different measurements, as the tracker enables the particle
momentum and charge determination, whereas the calorimeter is sensitive only to
the particle energy.  One should note that the preferable measurement is the
one with the tracker, as it provides very high precision and allows to
efficiently discriminate against the like-charge background pairs. Since the
correlation between the pseudorapidities of both pions is very large, the
following analysis is performed independently for the tracking detector
($|\eta|<2.5$) and the forward calorimetry ($2.5<|\eta|<4.9$).

\begin{figure}[t]
  \includegraphics[width=0.49\textwidth]{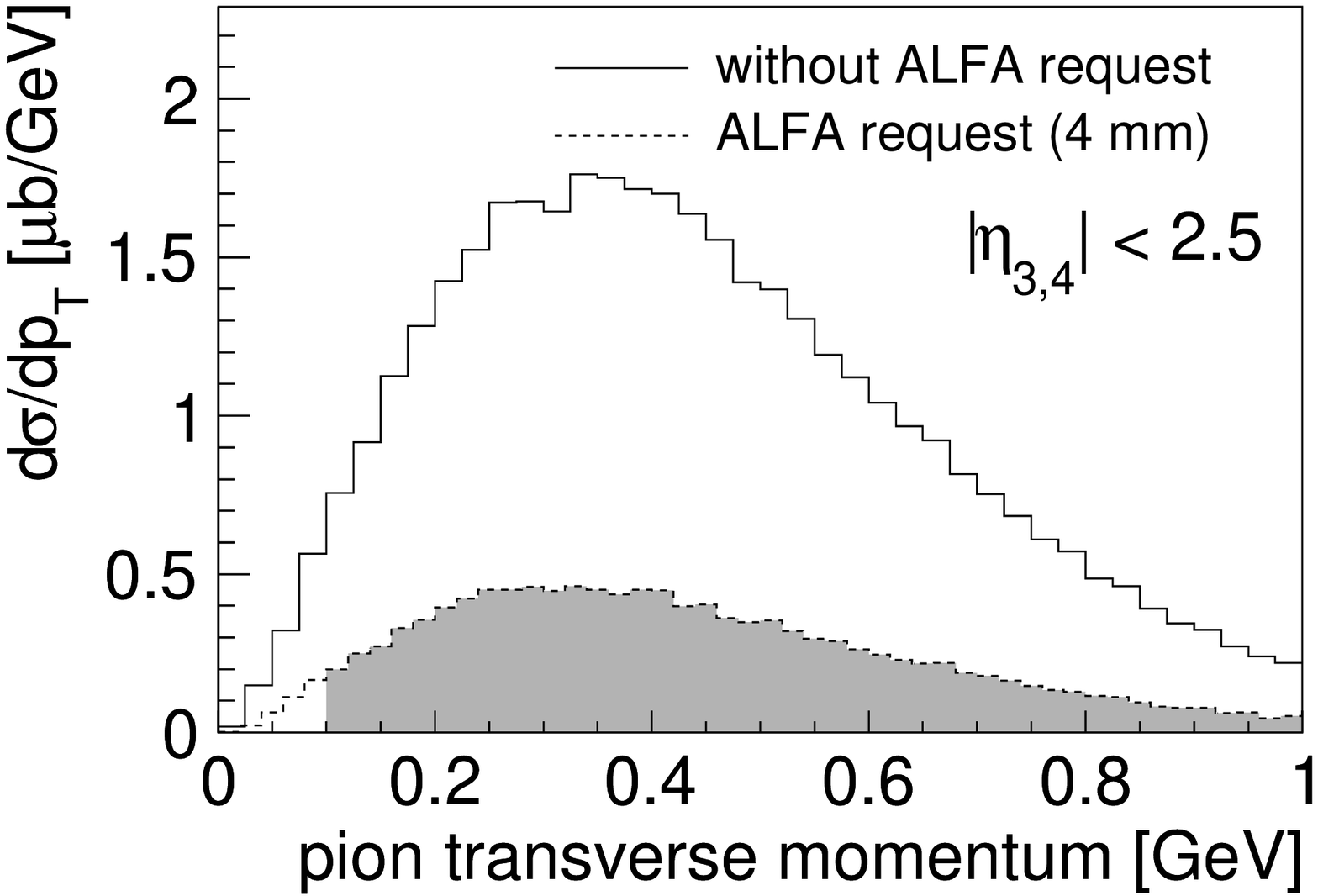}
  \hfill
  \includegraphics[width=0.49\textwidth]{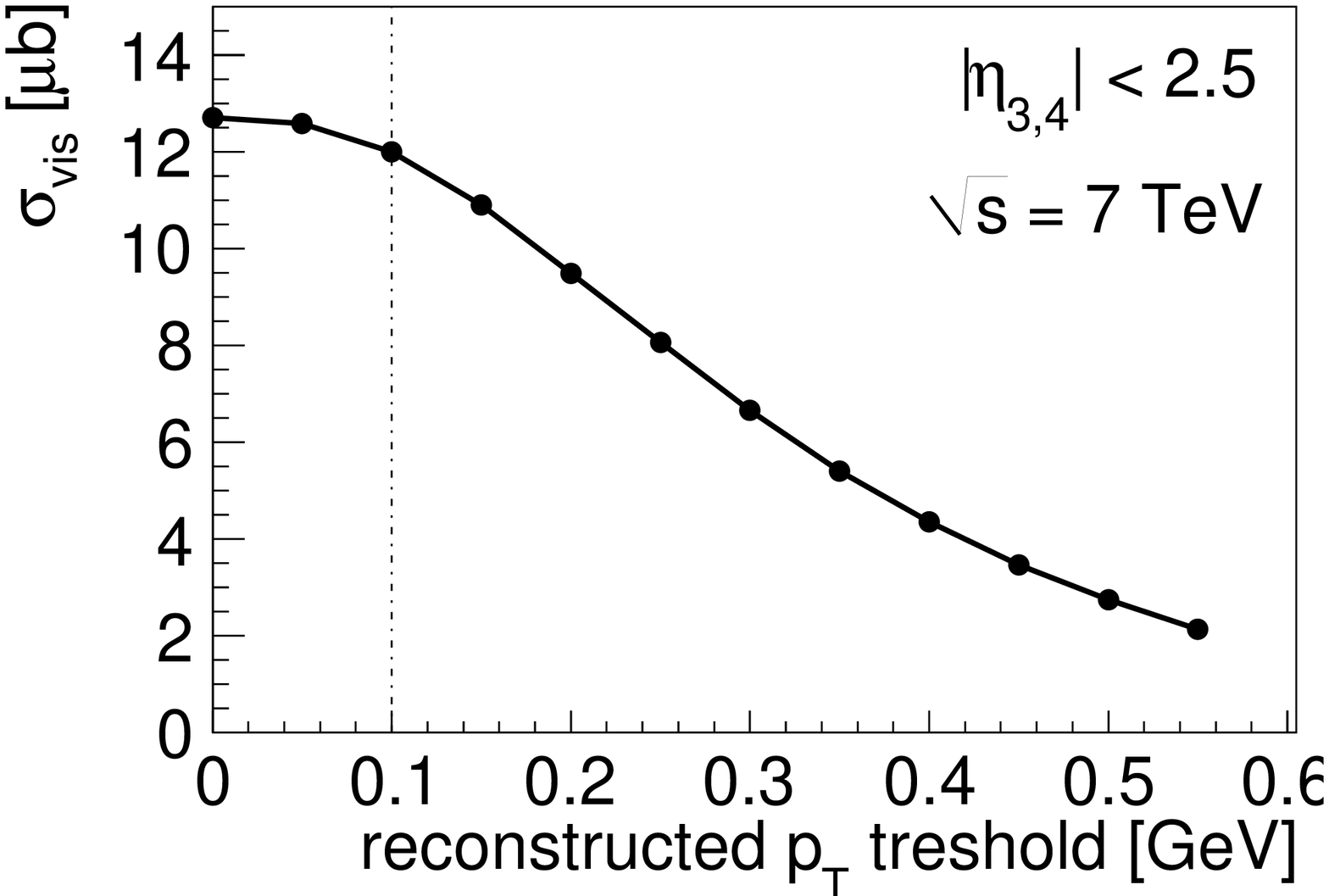}
  \caption{Left: the pion transverse momentum distribution in the tracking
  detector. Right: cross section for $|\eta|<2.5$ as a function of $p_T$
  treshold. The grey area and the dash-dotted line marks the lower boundary of the region accessible by
  ATLAS.}
  \label{fig:pt}
\end{figure}

\begin{figure}[t]
  \includegraphics[width=0.49\textwidth]{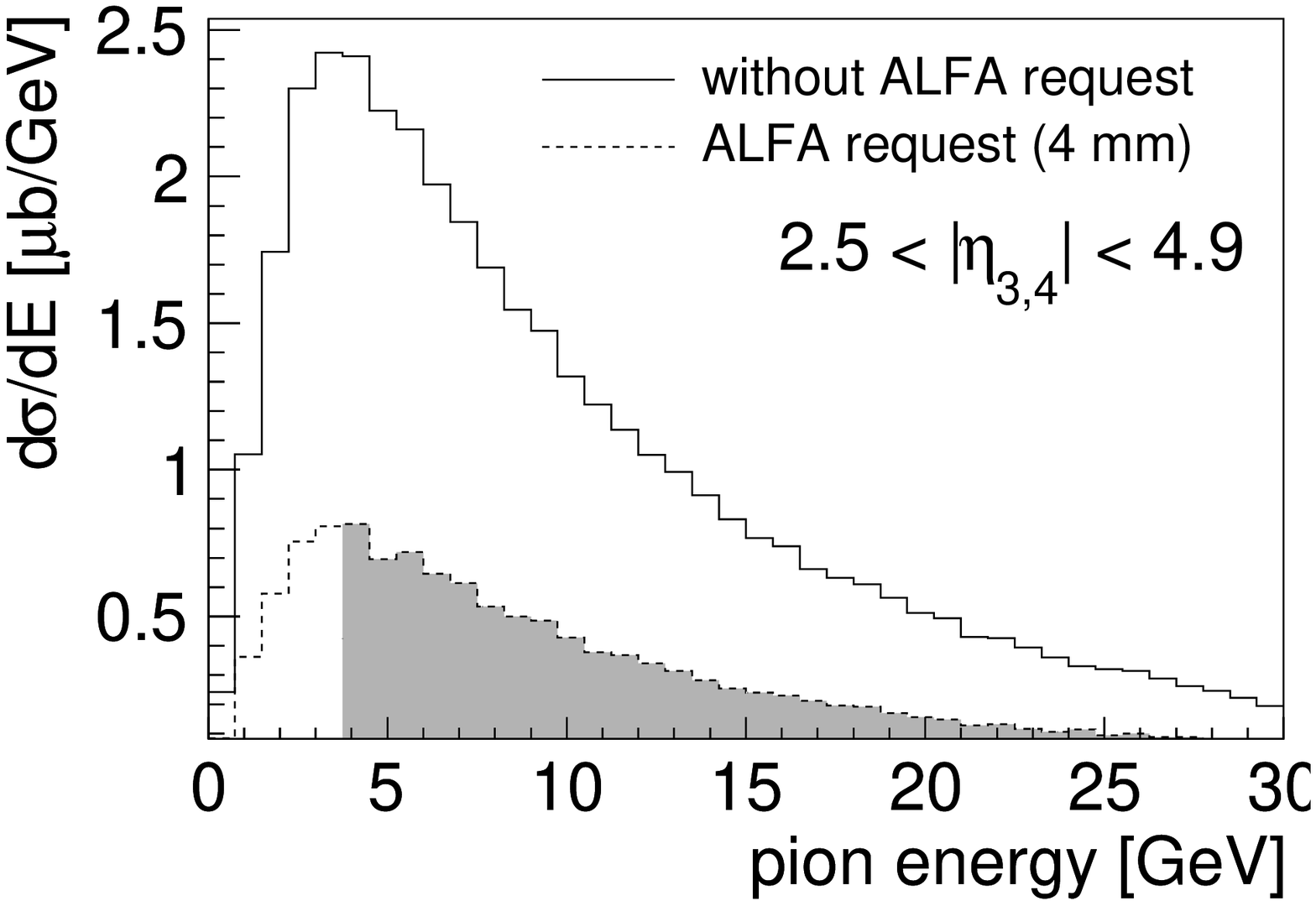}
  \hfill
  \includegraphics[width=0.49\textwidth]{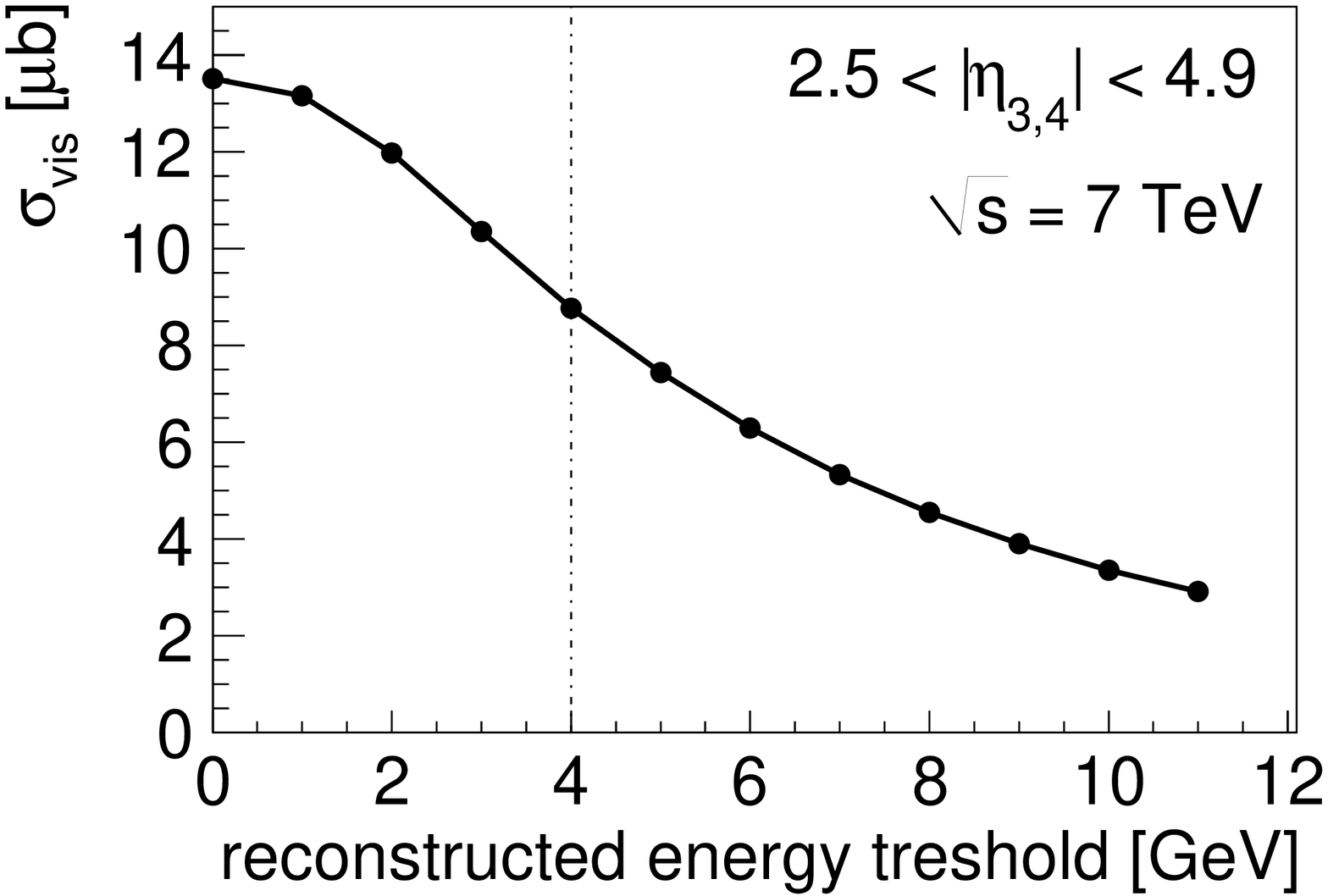}
  \caption{Left: the pion energy distribution in the calorimeter. Right: cross
  section for $4.9<|\eta|<2.5$ as a function of energy threshold. The grey area
  and the dash-dotted line marks the lower boundary of the region accessible by ATLAS.}
  \label{fig:E}
\end{figure}

The two adequate distributions: pion transverse momentum in the
central region ($|\eta|<2.5$) and pion energy in the forward region
($2.5<|\eta|<4.9$) are presented in Fig.~\ref{fig:pt} (left) and
Fig.~\ref{fig:E} (left). Requirement of both protons being tagged in the ALFA
detectors influences the shapes of the distributions only very little, but it
reduces both by a factor close to three.

Obviously, the number of events that can be observed depends on the minimal
pion transverse momentum and minimal pion energy that are experimentally
accessible.  Fig.~\ref{fig:pt} (right) and Fig.~\ref{fig:E} (right) show the
visible cross section as a function of reconstruction thresholds for measurements in
the tracker and the calorimeter.  Clearly, the cross section falls very steeply with
increasing thresholds values. The vertical dash-dotted lines show the
thresholds that should be possible to obtain: measurements of $p_T=100$~MeV
were performed for ATLAS minimum bias analysis \cite{min_bias} and particles
with energy $E > 4$~GeV were shown to be well above the noise \cite{FCAL}. It
should be mentioned that in the minimum bias analysis the efficiency for such
low-$p_T$ tracks was quite small (about 10\%). However, in that analysis, the
reconstruction algorithms had to simultaneously deal with many particle tracks.
For very clean events that are considered in this work (only two tracks) it
should be possible to adjust the reconstruction to obtain a much better
efficiency.

\begin{figure}[t]
  \includegraphics[width=0.49\textwidth]{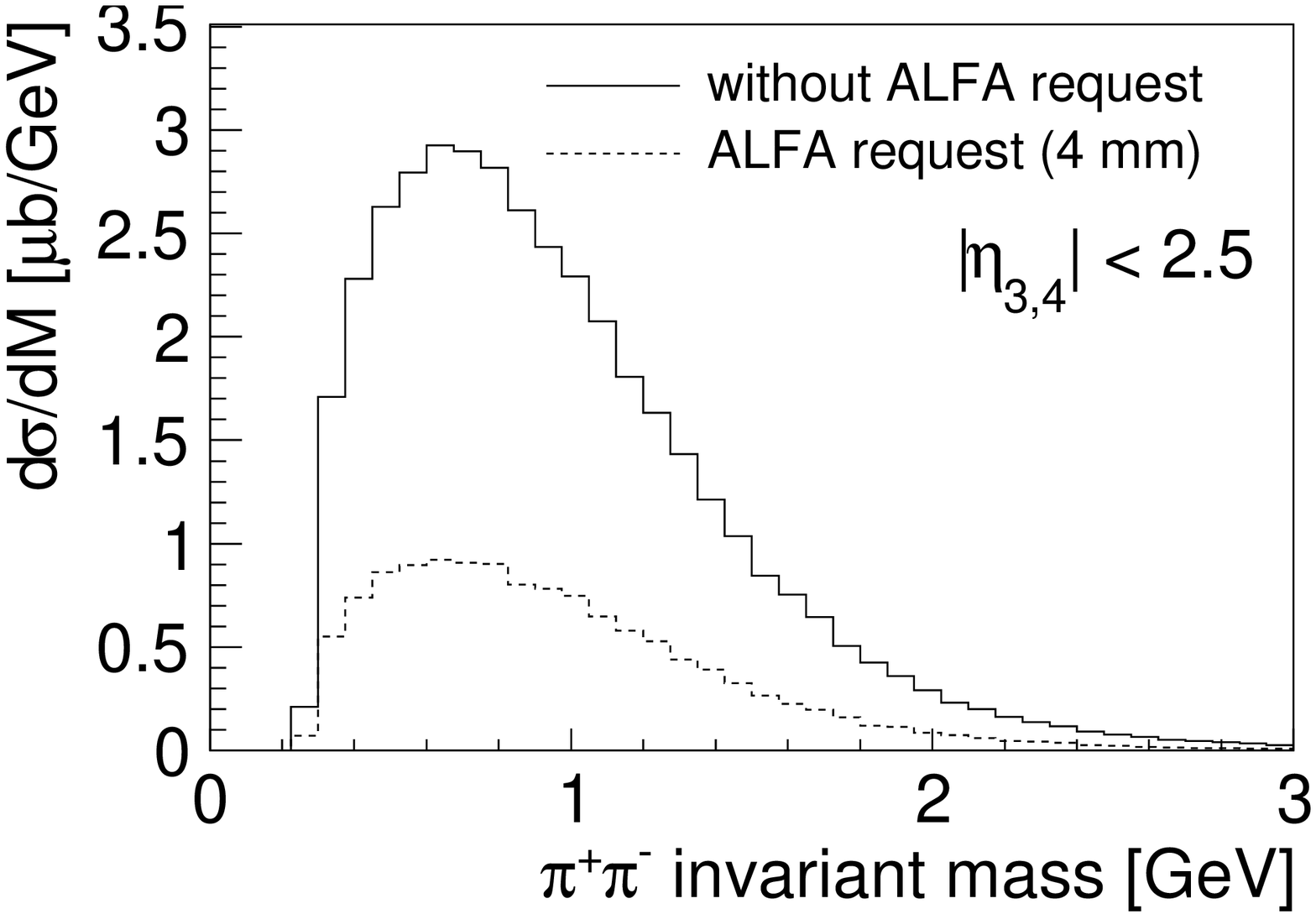}
  \hfill
  \includegraphics[width=0.49\textwidth]{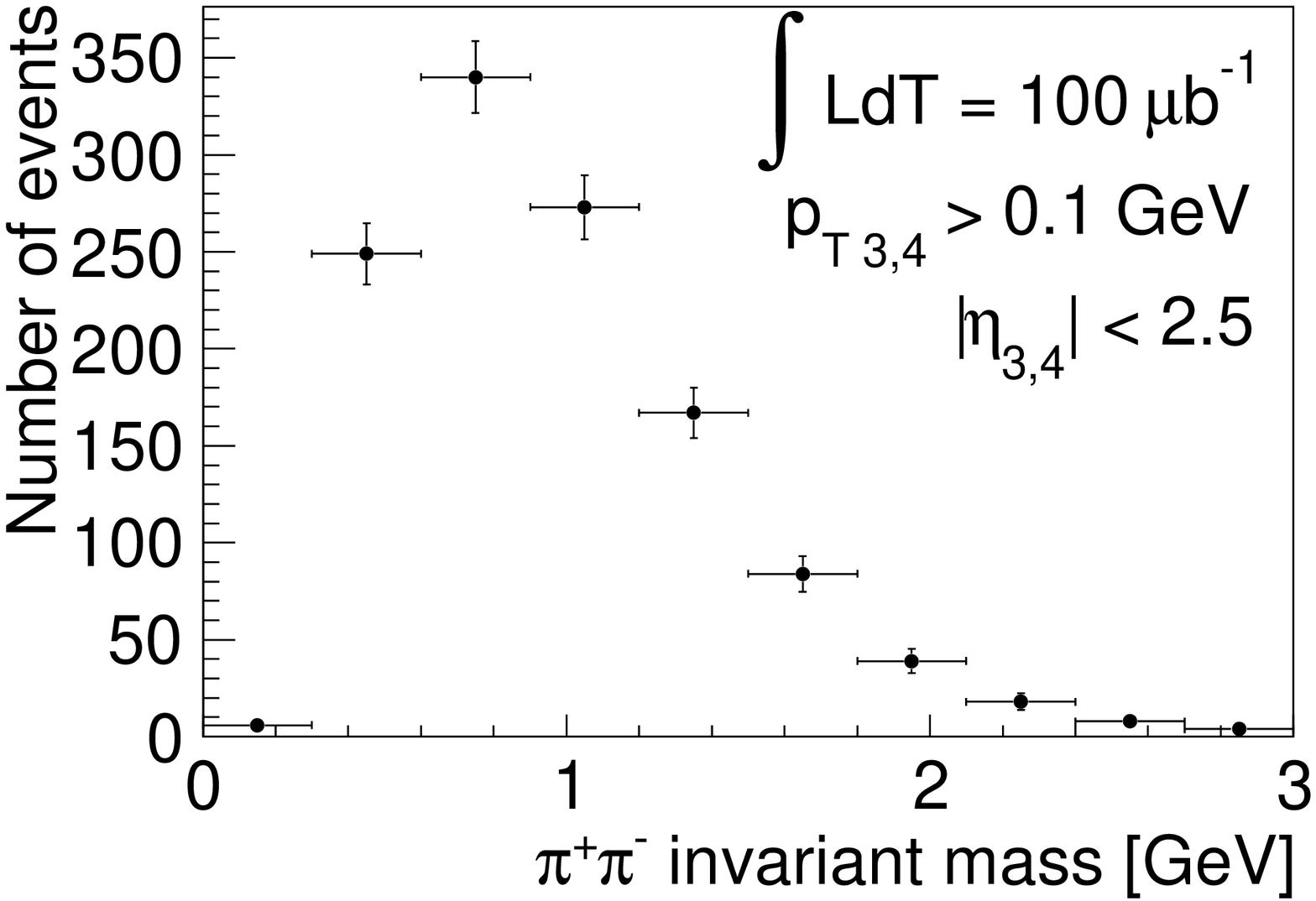}
  \caption{Left: distribution of $\pi^+\pi^-$ invariant mass reconstructed in
  the tracking detector. Right: possible measurement of the $\pi^+\pi^-$
  invariant mass distribution for 100~$\mu$b$^{-1}$ integrated luminosity (only
  the statistical errors are plotted).}
  \label{fig:M}
\end{figure}

An interesting study that can be made when data are collected is the measurement of
the $\pi^+\pi^-$ invariant mass distribution. Fig.~\ref{fig:M} presents the
theoretical predictions and a possible measurement with 100~$\mu$b$^{-1}$
integrated luminosity (30~hours of data acquisition time assuming the luminosity value of
$10^{27}$~cm$^{-2}$s$^{-1}$) for pions detected in the ATLAS tracker
(systematic uncertainty of such a measurement is not considered, only the
statistical errors are presented).  If the collected statistics is high enough,
it should be possible to see resonances, especially the $f_2(1270)$ meson, on top of the presented
distribution.

\section{Summary}

A process of exclusive pion pair production in proton-proton collisions was
presented and its theoretical description was briefly described. With this
theoretical model, a possibility of measuring this process at the LHC was
investigated for the ATLAS central detector and the ALFA very forward
detectors.  There are three main experimental parameters that limit the visible
cross section: the distance between the ALFA detector edge and the proton beam
centre, a minimal $p_T$ that can be measured in the tracking detector (for
pions produced in $|\eta|<2.5$) and minimal energy that can be measured in the
calorimeter (for $2.5<|\eta|<4.9$ range). For the values of these parameters
set to 4~mm, 0.1~GeV and 4~GeV, respectively, the visible cross section is
21~$\mu$b. For 100~$\mu$b$^{-1}$ of integrated luminosity that can be collected
during the ALFA runs this gives over 2000 events within the detector
acceptance.
\vspace{2mm}

Acknowledgements: R.S. would like to thank J.~Turnau for a discussion that
triggered this paper and D.~Derendarz for his comments about the experimental
aspects of the analysis.

\end{document}